\documentclass[12pt,letterpaper]{article}
\usepackage{epsfig,rotating,setspace,latexsym,amsmath,epsf,amssymb,bm,theorem}
\usepackage{cite,appendix}

\usepackage{amsfonts}

\title{Degrees of Freedom Region of the Gaussian MIMO Broadcast Channel with Common and Private Messages\thanks{This work
was supported by NSF Grants CCF 07-29127, CNS 09-64632, CCF
09-64645 and CCF 10-18185, and presented in part at the IEEE
Global Communications Conference, Miami, FL, December 2010.}}
\author{Ersen Ekrem \qquad Sennur Ulukus
\\
\normalsize Department of Electrical and Computer Engineering\\
\normalsize University of Maryland, College Park, MD 20742 \\
\normalsize {\it ersen@umd.edu} \qquad {\it ulukus@umd.edu}}

\newcommand{\bblambda}{\bm \Lambda}

\newcommand{\bbpsi}{\bm \Psi}
\newcommand{\bbomega}{\bm \Omega}

\newcommand{\bbsigma}{\bm \Sigma}

\newcommand{\bbi}{{\mathbf{I}}}
\newcommand{\bzero}{{\mathbf{0}}}

\newcommand{\bbh}{{\mathbf{H}}}

\newcommand{\bbk}{{\mathbf{K}}}

\newcommand{\bbn}{{\mathbf{N}}}

\newcommand{\bba}{{\mathbf{A}}}
\newcommand{\bbd}{{\mathbf{D}}}

\newcommand{\bbb}{{\mathbf{B}}}

\newcommand{\bbs}{{\mathbf{S}}}

\newcommand{\bbx}{{\mathbf{X}}}

\newcommand{\bby}{{\mathbf{Y}}}

\newtheorem{Theo}{Theorem}

\newtheorem{Def}{Definition}

\begin{document}


\maketitle

\begin{abstract}
We consider the Gaussian multiple-input multiple-output (MIMO)
broadcast channel with common and private messages. We obtain the
degrees of freedom (DoF) region of this channel. We first show
that a parallel Gaussian broadcast channel with unmatched
sub-channels can be constructed from any given Gaussian MIMO
broadcast channel by using the generalized singular value
decomposition (GSVD) and a relaxation on the power constraint for
the channel input, in a way that the capacity region of the
constructed parallel channel provides an outer bound for the
capacity region of the original channel. The capacity region of
the parallel Gaussian broadcast channel with unmatched
sub-channels is known, using which we obtain an explicit outer
bound for the DoF region of the Gaussian MIMO broadcast channel.
We finally show that this outer bound for the DoF region can be
attained both by the achievable scheme that uses a classical
Gaussian coding for the common message and dirty-paper coding
(DPC) for the private messages, as well as by a variation of the
zero-forcing (ZF) scheme.
\end{abstract}

\newpage

\section{Introduction}

We study the two-user Gaussian multiple-input multiple-output
(MIMO) broadcast channel, where each link between the transmitter
and each receiver is a linear additive Gaussian channel. We
consider the scenario where the transmitter sends a private
message to each user in addition to a common message which is
directed to both users. The capacity region for this scenario,
i.e., the capacity region of the Gaussian MIMO broadcast channel
with common and private messages, is unknown. However, when one of
these three messages is absent, the corresponding capacity region
is known. In particular, the capacity region is known when there
is no common message, i.e., each user gets only a private
message~\cite{Shamai_MIMO}, and for the degraded message set case,
i.e., there is a common message directed to both users, and only
one of the users gets a private
message~\cite{Hannan_Common,hannan_thesis}.

The first work that considers the Gaussian MIMO broadcast channel
with common and private messages is~\cite{Goldsmith_common}.
Reference~\cite{Goldsmith_common} proposes an achievable scheme
which uses a classical Gaussian coding scheme for the common
message, and dirty-paper coding (DPC) for the private messages.
The corresponding achievable rate region is called the DPC region.
In addition,~\cite{Goldsmith_common} obtains the capacity region
when the Gaussian MIMO broadcast channel is equivalent to a set of
parallel independent Gaussian channels by using the results
from~\cite{El_Gamal_Product}. The Gaussian MIMO broadcast channel
with common and private messages is further studied
in~\cite{Hannan_Common, hannan_thesis}, where the partial
optimality of the DPC region~\cite{Goldsmith_common} is shown.
References~\cite{Hannan_Common, hannan_thesis} first propose an
outer bound for the capacity region of the Gaussian MIMO broadcast
channel with common and private messages, and then prove that it
is tight on certain sub-regions of the capacity region by showing
that it matches the DPC region given in~\cite{Goldsmith_common}.
Moreover, \cite{Hannan_Common,hannan_thesis} show that for a given
common message rate, the private message sum capacity is attained
by the achievable scheme in~\cite{Goldsmith_common}. Finally,
\cite{Hannan_Common,hannan_thesis} show the optimality of the DPC
region in~\cite{Goldsmith_common} when the common message rate is
beyond a certain threshold.

A more recent work on the Gaussian MIMO broadcast channel is
reported in~\cite{MIMO_Common_Private}\footnote{Some of the
results in~\cite{MIMO_Common_Private} are concurrently and
independently obtained in~\cite{MIMO_Common_Private_Shamai}.}.
In~\cite{MIMO_Common_Private}, we first obtain an outer bound for
the capacity region of the two-user discrete memoryless broadcast
channel with common and private messages. We next show that if
jointly Gaussian random variables are sufficient to evaluate this
outer bound for the Gaussian MIMO broadcast channel, the DPC
region is the capacity region of the Gaussian MIMO broadcast
channel with common and private messages. However, we can evaluate
only a loosened version of this outer bound, which yields the
result that extending the DPC region in the common message rate
direction by a fixed amount is an outer bound for the capacity
region of the Gaussian MIMO broadcast channel with common and
private messages. However, this fixed amount, i.e., the gap, does
not have suitable scaling with the available power at the
transmitter to enable us to obtain the degrees of freedom (DoF)
region of the Gaussian MIMO broadcast channel with common and
private messages.

In this work, we follow a different approach and establish the DoF
region of the Gaussian MIMO broadcast channel with common and
private messages. We first show that we can construct a parallel
Gaussian broadcast channel with unmatched
sub-channels~\cite{El_Gamal_Product} from any given Gaussian MIMO
broadcast channel such that the capacity region of this parallel
Gaussian broadcast channel with unmatched sub-channels includes
the capacity region of the Gaussian MIMO broadcast channel. To
construct such a parallel channel, we use the generalized singular
value decomposition (GSVD)~\cite{GSVD} on the channel gain
matrices of the Gaussian MIMO broadcast channel and also relax the
power constraint on the channel input. This relaxation on the
power constraint enlarges the capacity region during the
transformation of the Gaussian MIMO broadcast channel into a
parallel Gaussian broadcast channel with unmatched sub-channels.
Consequently, the capacity region of the constructed parallel
channel provides an outer bound for the capacity region of the
Gaussian MIMO channel. Since the capacity region of the parallel
Gaussian broadcast channel with unmatched sub-channels is known
due to~\cite{El_Gamal_Product}, we are able to characterize the
DoF region of the parallel Gaussian broadcast channel with
unmatched sub-channels, which serves as an outer bound for the DoF
region of the Gaussian MIMO broadcast channel. We next show that
this outer bound for the DoF region of the Gaussian MIMO broadcast
channel with common and private messages can be attained by a
proper selection of the covariance matrices involved in the DPC
region~\cite{Goldsmith_common}. Moreover, we also show that, in
addition to the DPC scheme, a variation of the zero-forcing (ZF)
scheme~\cite{zf_1,zf_2} can attain the DoF region of the Gaussian
MIMO broadcast channel with common and private messages.

\section{Channel Model and Definitions}

The Gaussian MIMO broadcast channel is defined by
\begin{align}
\bby_1&=\bbh_1 \bbx+\bbn_1 \label{general_gaussian_mimo_1}\\
\bby_2&=\bbh_2 \bbx+\bbn_2 \label{general_gaussian_mimo_2}
\end{align}
where the channel input $\bbx$ is a $t\times 1$ column vector,
$\bbh_j$ is the $j$th user's channel gain matrix of size
$r_j\times t$, $\bby_j$ is the channel output of the $j$th user
which is an $r_j\times 1$ column vector, and the Gaussian random
vector $\bbn_j$ is of size $r_j\times 1$ with an identity
covariance matrix. The channel input is subject to an average
power constraint as follows
\begin{align}
E\left[\bbx^{\top}\bbx\right]={\rm tr}\left(E\left[\bbx
\bbx^{\top}\right]\right)\leq P \label{trace_constraint}
\end{align}

We study the Gaussian MIMO broadcast channel for the scenario
where the transmitter sends a common message to both users, and a
private message to each user. We call the channel model arising
from this scenario the {\em Gaussian MIMO broadcast channel with
common and private messages}. An $(n,2^{nR_0},2^{nR_1},2^{nR_2})$
code for this channel consists of three message sets
$\mathcal{W}_0=\{1,\ldots,
2^{nR_0}\},\mathcal{W}_1=\{1,\ldots,2^{nR_1}\},
\mathcal{W}_2=\{1,\ldots,2^{nR_2}\}$, one encoder
$f_n:\mathcal{W}_0\times\mathcal{W}_1\times
\mathcal{W}_2\rightarrow \mathcal{X}^n$, one decoder at each
receiver $g_n^j:\mathcal{Y}_j^n\rightarrow
\mathcal{W}_0\times\mathcal{W}_j,~j=1,2.$ The probability of error
is defined as $P_e^n=\max\{P_{e1}^n,P_{e2}^n\}$, where
$P_{ej}=\Pr[g_n^j(f_n(W_0,W_1,W_2))\neq (W_0,W_j)],~j=1,2$, and
$W_j$ denotes the message which is a uniformly distributed random
variable in $\mathcal{W}_j,~j=0,1,2$. A rate triple
$(R_0,R_1,R_2)$ is said to be achievable if there exists a code
$(n,2^{nR_0},2^{nR_1},2^{nR_2})$ which has $\lim_{n\rightarrow
\infty}P_e^n=0$. The capacity region $\mathcal{C}(P)$ is defined
as the convex closure of all achievable rate triples
$(R_0,R_1,R_2)$.

Our main concern is to investigate how the capacity region
$\mathcal{C}(P)$ behaves when the available power at the
transmitter $P$ is arbitrarily large, i.e., $P$ goes to infinity.
This investigation can be carried out by characterizing the DoF
region of the Gaussian MIMO broadcast channel with common and
private messages. A DoF triple $(d_0,d_1,d_2)$ is said to be
achievable if there exists a rate triple
$(R_0,R_1,R_2)\in\mathcal{C}(P)$ such that
\begin{align}
d_j&=\lim_{P\rightarrow \infty} \frac{R_j}{\frac{1}{2}\log
P},\quad j=0,1,2
\end{align}
The DoF region $\mathcal{D}$ is defined as the convex closure of
all achievable DoF triples $(d_0,d_1,d_2)$.

We conclude this section by presenting the achievable rate region,
hereafter called the DPC region, given in~\cite{Goldsmith_common}.
In the achievable scheme in~\cite{Goldsmith_common}, the common
message is encoded by a standard Gaussian codebook, and the
private messages are encoded by DPC. Each user decodes the common
message by treating the signals carrying the private messages as
noise. Next, users decode their private messages. Since a DPC
scheme is used to encode the private messages, one of the users
observes an interference-free link depending on the encoding order
at the transmitter. We next define
\begin{align}
\hspace{-0.1cm}R_{0j}(\bbk_0,\bbk_1,\bbk_2)&=
\frac{1}{2}\log\frac{|\bbh_j(\bbk_0+\bbk_1+\bbk_2)\bbh_j^{\top}+\bbi|}{|\bbh_j(\bbk_1+\bbk_2)\bbh_j^{\top}
+\bbi|},\quad j=1,2 \label{common_message_rate}\\
R_1(\bbk_1,\bbk_2)&= \frac{1}{2}
\log\frac{|\bbh_1(\bbk_1+\bbk_2)\bbh_1^{\top}+\bbi|}{|\bbh_1\bbk_2\bbh_1^{\top}+\bbi|} \label{private_message_1_rate}\\
R_2(\bbk_2)&= \frac{1}{2} \log|\bbh_2\bbk_2
\bbh_2+\bbi|\label{private_message_2_rate}
\end{align}
where $\bbk_0,\bbk_1,\bbk_2$ denote the covariance matrices
allotted for the common message, the first user's private message,
and the second user's private message, respectively. The DPC
region is stated in the following theorem.
\begin{Theo}[\!\!\cite{Goldsmith_common}]
\vspace{-0.4cm} \label{proposition_ach} The rate triples
$(R_0,R_1,R_2)$ lying in the region
\begin{align}
\mathcal{R}^{\rm DPC}(P)={\rm conv}\left(\mathcal{R}_1^{\rm
DPC}(P)\cup \mathcal{R}_2^{\rm DPC}(P)\right)
\end{align}
are achievable, where ${\rm conv}$ is the convex hull operator,
$\mathcal{R}_1^{\rm DPC}(P)$ consists of rate triples
$(R_0,R_1,R_2)$ satisfying
\begin{align}
R_0 &\leq R_{0j}(\bbk_0,\bbk_1,\bbk_2),\quad j=1,2\\
R_1 &\leq R_1(\bbk_1,\bbk_2)\\
R_2&\leq R_2(\bbk_2)
\end{align}
for some positive semi-definite matrices $\bbk_0,\bbk_1,\bbk_2$
such that ${\rm tr}(\bbk_0+\bbk_1+\bbk_2)\leq P$, and
$\mathcal{R}_2^{\rm DPC}(P)$ can be obtained from
$\mathcal{R}_1^{\rm DPC}(P)$ by swapping the subscripts 1 and 2.
\vspace{-0.25cm}
\end{Theo}

The DPC region is tight in several cases. The first one is the
case where each receiver gets only a private message, i.e.,
$R_0=0$~\cite{Shamai_MIMO}. The other case is the degraded message
sets scenario in which we have either $R_1=0$ or
$R_2=0$~\cite{Hannan_Common}. In both of these cases, there are
only two messages to be sent. The case when both private messages
and a common message are present is investigated
in~\cite{Hannan_Common,hannan_thesis}.
In~\cite{Hannan_Common,hannan_thesis}, outer bounds on the
capacity region with private and common messages are given and
these outer bounds are shown to match the DPC region in certain
regions. Furthermore, \cite{Hannan_Common,hannan_thesis} show that
for a given common message rate $R_0$, the DPC region achieves the
private message sum rate capacity, i.e., the maximum of $R_1+R_2$.
Finally, \cite{Hannan_Common,hannan_thesis} show that if the
common message rate is beyond a certain threshold, the DPC region
matches the capacity region if the channel input is subject to a
covariance constraint, i.e., $E\left[\bbx\bbx^\top\right]\preceq
\bbs$ for some $\bbs\succeq \bzero$.
In~\cite{MIMO_Common_Private}, we show that an outer bound for the
capacity region of the Gaussian MIMO broadcast channel with common
and private messages can be obtained by extending the DPC region
in the common message rate direction by a fixed amount. This fixed
amount, i.e., the gap, depends on the channel gain matrices
$\bbh_1,\bbh_2$, and is not finite for all possible channel gain
matrices $\bbh_1,\bbh_2$.

\section{Main Result}
We now present our main result which characterizes the DoF region
of the Gaussian MIMO broadcast channel with common and private
messages. Our result shows that this DoF region can be attained by
using the achievable scheme in Theorem~\ref{proposition_ach},
i.e., the DPC region in Theorem~\ref{proposition_ach} is
asymptotically tight. Moreover, we also show that in addition to
the achievable scheme in Theorem~\ref{proposition_ach}, a
variation of the ZF scheme~\cite{zf_1,zf_2} can achieve the DoF
region as well. Before stating our main result, we introduce the
GSVD~\cite{GSVD,Wornell} which plays a crucial role in the proof
of our main result, and provides the necessary notation to express
this result.

\begin{Def}[\!\!\cite{GSVD}, Theorem~1]
Given two matrices $\bbh_1\in\mathbb{R}^{r_1\times t}$ and
$\bbh_2\in\mathbb{R}^{r_2\times t}$, there exist orthonormal
matrices $\bbpsi_1\in\mathbb{R}^{r_1\times
r_1},\bbpsi_2\in\mathbb{R}^{r_2\times
r_2},\bbpsi_0\in\mathbb{R}^{t\times t}$, a non-singular, lower
triangular matrix $\bbomega\in\mathbb{R}^{k\times k}$, and two
matrices $\bbsigma_1 \in\mathbb{R}^{r_1\times k},\bbsigma_2
\in\mathbb{R}^{r_2\times k}$ such that
\begin{align}
\bbpsi_1^\top\bbh_1\bbpsi_0&=\bbsigma_1 \left[~\bbomega^{-1}~~
\bzero_{k\times t-k}~\right]\label{decompose_h1}\\
\bbpsi_2^\top \bbh_2\bbpsi_0&=\bbsigma_2 \left[~\bbomega^{-1}~~
\bzero_{k\times t-k}~\right] \label{decompose_h2}
\end{align}
where $\bbsigma_1$ and $\bbsigma_2$ are given by
\begin{align}
\bbsigma_1&=\left[
\begin{array}{rcl}
\bbi_{k-p-s\times k-p-s} && \\
&\bbd_{1,s\times s}& \\
&& \bzero_{r_1+p-k\times p }
\end{array}
\right] \\
\bbsigma_2&=\left[
\begin{array}{rcl}
\bzero_{r_2-p-s\times k-p-s} && \\
&\bbd_{2,s\times s}& \\
&& \bbi_{p\times p }
\end{array}
\right]
\end{align}
and the constants $k,p$ are given as
\begin{align}
k&={\rm rank} \left(\left[
\begin{array}{rcl}
\bbh_1\\
\bbh_2
\end{array}
\right]\right)\\
p&={\rm dim}\left({\rm Null}(\bbh_1)\cap {\rm Null}(\bbh_2)^\bot
\right)
\end{align}
and s depends on the matrices $\bbh_1,\bbh_2$. The matrices
$\bbd_1,\bbd_2$ are diagonal with the diagonal elements being
strictly positive.
\end{Def}

We define the sets $\mathcal{S}_1,\mathcal{S}_c,\mathcal{S}_2$ as
follows
\begin{align}
\mathcal{S}_1&=\{1,\ldots,k-p-s\} \label{set_1} \\
\mathcal{S}_c&=\{k-p-s+1,\ldots,k-p\} \label{set_common}\\
\mathcal{S}_2&=\{k-p+1,\ldots,k\} \label{set_2}
\end{align}
Our main result is stated in the following theorem.
\begin{Theo}
\label{theorem_dof} The DoF region of the Gaussian MIMO broadcast
channel with common and private messages is given by the union of
DoF triples $(d_0,d_1,d_2)$ satisfying
\begin{align}
d_0&\leq |\mathcal{S}_c|-\alpha_1-\alpha_2+\beta \\
d_1&\leq \alpha_1+|\mathcal{S}_1|-\beta \\
d_2&\leq \alpha_2+|\mathcal{S}_2|-\beta
\end{align}
for some non-negative $\alpha_1,\alpha_2,\beta$ such that
$\alpha_1+\alpha_2 \leq |\mathcal{S}_c|$, $\beta\leq
\min\{|\mathcal{S}_1|,|\mathcal{S}_2|\}$. The DoF region of the
Gaussian MIMO broadcast channel with common and private messages
can be attained by the DPC region given in
Theorem~\ref{proposition_ach} as well as by a variation of the ZF
scheme.
\end{Theo}

This theorem states that, if the available power $P$ is
sufficiently large, the Gaussian MIMO broadcast channel behaves as
if it is a parallel Gaussian broadcast channel with
$|\mathcal{S}_1|+|\mathcal{S}_c|+|\mathcal{S}_2|$ sub-channels.
$|\mathcal{S}_1 |$ of these sub-channels can be accessed by only
the first user, $|\mathcal{S}_2|$ of these sub-channels can be
accessed by only the second user, and $|\mathcal{S}_c|$ of these
sub-channels can be accessed by both users. For a fixed
$(\alpha_1,\alpha_2,\beta)$, $|\mathcal{S}_c|-\alpha_1-\alpha_2$
of the sub-channels that both users can access, need to be used
for the transmission of the common message in addition to $\beta$
of the $|\mathcal{S}_1|$ sub-channels that only the first user can
access and $\beta$ of the $|\mathcal{S}_2|$ sub-channels that only
the second user can access. Thus, each user gets the common
message over some common sub-channels, namely
$|\mathcal{S}_c|-\alpha_1-\alpha_2$ sub-channels, which can be
observed by both users, and some private sub-channels, namely
$\beta$ sub-channels, which can be observed by only one user. This
leads to a $\beta+|\mathcal{S}_c|-\alpha_1-\alpha_2$ DoF for the
common message. The first user's message needs to be transmitted
over $\alpha_1$ of the $|\mathcal{S}_c|$ sub-channels that are
observed by both users, and the remaining $|\mathcal{S}_1|-\beta$
of the first user's private sub-channels (the rest of these
sub-channels were dedicated to the transmission of the common
message) that cannot be observed by the second user. This results
in an $\alpha_1+|\mathcal{S}_1|-\beta$ DoF for the first user's
private message. Similarly, the second user's message needs to be
transmitted over $\alpha_2$ of the $|\mathcal{S}_c|$ sub-channels
that are observed by both users, and the remaining
$|\mathcal{S}_2|-\beta$ of the second user's private sub-channels
(the rest of these sub-channels were dedicated to the transmission
of the common message) that cannot be observed by the first user.
This results in an $\alpha_2+|\mathcal{S}_2|-\beta$ DoF for the
second user's private message.

We provide the proof of Theorem~\ref{theorem_dof} in the next
three sections. In the next section, we obtain an outer bound for
the DoF region of the Gaussian MIMO broadcast channel with common
and private messages by using the GSVD, and also a relaxation on
the power constraint for the channel input. In
Section~\ref{sec:inner_bound}, we obtain an inner bound for the
DoF region of the Gaussian MIMO broadcast channel with common and
private messages. We obtain this inner bound by using two
different achievable schemes. The first one directly uses
Theorem~\ref{proposition_ach}, i.e., we make an explicit selection
of the covariance matrices $\bbk_0,\bbk_1,\bbk_2$ involved in the
DPC region to obtain this inner bound. The second one employs a
variation of the ZF scheme~\cite{zf_1,zf_2} to obtain the inner
bound for the DoF region. The equivalence of these inner and outer
bounds are shown in Section~\ref{sec:equivalence} to complete the
proof of Theorem~\ref{theorem_dof}.

\section{Outer Bound}
\label{sec:outer_bound}

We first obtain a new channel from the original one in
(\ref{general_gaussian_mimo_1})-(\ref{general_gaussian_mimo_2})-(\ref{trace_constraint})
by using the GSVD, where the capacity region of the new channel
includes the capacity region of the original one in
(\ref{general_gaussian_mimo_1})-(\ref{general_gaussian_mimo_2})-(\ref{trace_constraint}).
To this end, we note that
\begin{align}
\bbpsi_j^{\top}\bbh_j&=\bbsigma_j
\left[\bbomega^{-1}~~\bzero_{k\times
t-k}\right]\bbpsi_0^{\top},\quad j=1,2\label{gsvd_implies}
\end{align}
which is due to (\ref{decompose_h1})-(\ref{decompose_h2}), and the
fact that $\bbpsi_0$ is orthonormal. Since $\bbpsi_j$ is also
orthonormal, i.e., non-singular, the capacity region of the
following channel
\begin{align}
\tilde{\bby}_j=\bbpsi_j^\top\bby_j,\quad j=1,2
\label{intermediate_channel_1}
\end{align}
is equal to the capacity region of the original one in
(\ref{general_gaussian_mimo_1})-(\ref{general_gaussian_mimo_2})-(\ref{trace_constraint}).
The channel defined in (\ref{intermediate_channel_1}) can be
explicitly expressed as
\begin{align}
\tilde{\bby}_j&=\bbpsi_j^\top\bbh_j\bbx+\bbpsi_j^\top\bbn_j\\
&=\bbsigma_j \left[\bbomega^{-1}~~\bzero_{k\times
t-k}\right]\bbpsi_0^{\top}\bbx+\bbpsi_j^\top\bbn_j ,\quad j=1,2
\label{intermediate_channel_2}
\end{align}
where we used (\ref{gsvd_implies}). We define
$\tilde{\bbn}_j=\bbpsi_j^\top\bbn_j$ which is also a white
Gaussian random vector, i.e.,
\begin{align}
E\left[\tilde{\bbn}_j\tilde{\bbn}_j^\top\right]=\bbi
\label{noise_still_white}
\end{align}
due to the fact that $\bbpsi_j$ is orthonormal and $\bbn_j$ is
white. We also define
\begin{align}
\tilde{\bbx}=\left[\bbomega^{-1}~~\bzero_{k\times
t-k}\right]\bbpsi_0^{\top}\bbx \label{new_channel_input}
\end{align}
using which the channel
in (\ref{intermediate_channel_2}) can be written as
\begin{align}
\tilde{\bby}_j=\bbsigma_j\tilde{\bbx}+\tilde{\bbn}_j,\quad j=1,2
\label{towards_a_new_channel}
\end{align}
where the channel input $\tilde{\bbx}$ should be chosen according
to the trace constraint on $\bbx$ stated in
(\ref{trace_constraint}). We now relax the power constraint on
$\tilde{\bbx}$, and consequently, obtain a new channel whose
capacity region includes the capacity region of the original
channel in
(\ref{general_gaussian_mimo_1})-(\ref{general_gaussian_mimo_2})-(\ref{trace_constraint}).
To this end, we note that
\begin{align}
{\rm
tr}\left(E\left[\tilde{\bbx}\tilde{\bbx}^\top\right]\right)&={\rm
tr}\left(\left[\bbomega^{-1}~~\bzero_{k\times
t-k}\right]\bbpsi_0^{\top}
E\left[\bbx\bbx^\top\right]\bbpsi_0\left[\bbomega^{-1}~~\bzero_{k\times
t-k}\right]^\top\right) \label{def_tilde_x} \\
&={\rm tr}\left(
E\left[\bbx\bbx^\top\right]\bbpsi_0\left[\bbomega^{-1}~~\bzero_{k\times
t-k}\right]^\top \left[\bbomega^{-1}~~\bzero_{k\times
t-k}\right]\bbpsi_0^{\top}\right) \label{trace_exchange}
\end{align}
where (\ref{def_tilde_x}) comes from the definition of
$\tilde{\bbx}$ in (\ref{new_channel_input}), and
(\ref{trace_exchange}) comes from the fact that ${\rm
tr}(\bba\bbb)={\rm tr}(\bbb \bba)$. Since
\begin{align}
\left[\bbomega^{-1}~~\bzero_{k\times t-k}\right]^\top
\left[\bbomega^{-1}~~\bzero_{k\times t-k}\right]
\end{align}
is a positive semi-definite matrix, there exists a $\zeta> 0 $
such that
\begin{align}
\left[\bbomega^{-1}~~\bzero_{k\times t-k}\right]^\top
\left[\bbomega^{-1}~~\bzero_{k\times t-k}\right] \preceq \zeta
\bbi \label{relaxation}
\end{align}
Since ${\rm tr}(\bba \bbb)\geq 0$ if $\bba \succeq
\bzero,\bbb\succeq \bzero$, using (\ref{relaxation}) in
(\ref{trace_exchange}), we get
\begin{align}
{\rm
tr}\left(E\left[\tilde{\bbx}\tilde{\bbx}^\top\right]\right)&\leq
\zeta {\rm tr}\left(
E\left[\bbx\bbx^\top\right]\bbpsi_0\bbpsi_0^{\top}\right)
\\
&= \zeta {\rm tr}\left(
E\left[\bbx\bbx^\top\right]\right)\label{orthonormality}
\\
&\leq \zeta P \label{trace_constraint_implies}
\end{align}
where (\ref{orthonormality}) comes from the fact that $\bbpsi_0$
is orthonormal, and (\ref{trace_constraint_implies}) is due to the
total power constraint on $\bbx$ given in
(\ref{trace_constraint}). We now consider the following channel
\begin{align}
\tilde{\bby}_j&=\bbsigma_j\tilde{\bbx}+\tilde{\bbn}_j,\quad j=1,2
\label{new_channel}
\end{align}
where the channel input is subject to the following trace
constraint
\begin{align}
{\rm tr}\left(E\left[\tilde{\bbx}\tilde{\bbx}^\top\right]\right)
&\leq \zeta P \label{new_trace_constraint}
\end{align}
We note that this new channel in
(\ref{new_channel})-(\ref{new_trace_constraint}) is obtained from
the original channel in
(\ref{general_gaussian_mimo_1})-(\ref{general_gaussian_mimo_2})-(\ref{trace_constraint})
by two main operations: The first one is the multiplication of the
channel outputs in the original channel, i.e.,
(\ref{general_gaussian_mimo_1})-(\ref{general_gaussian_mimo_2}),
with invertible matrices $\bbpsi_1,\bbpsi_2$ which preserves the
capacity region. The second operation is the relaxation of the
power constraint in the new channel to get
(\ref{new_trace_constraint}) which increases the capacity region
by means of increasing the set of all feasible input
distributions. Thus, due to this second operation, the capacity
region of the new channel in
(\ref{new_channel})-(\ref{new_trace_constraint}) serves as an
outer bound for the capacity region of the original channel in
(\ref{general_gaussian_mimo_1})-(\ref{general_gaussian_mimo_2})-(\ref{trace_constraint}).
Similarly, the DoF region of the new channel in
(\ref{new_channel})-(\ref{new_trace_constraint}) is an outer bound
for the DoF region of the original channel in
(\ref{general_gaussian_mimo_1})-(\ref{general_gaussian_mimo_2})-(\ref{trace_constraint}).

We next rewrite the channel in
(\ref{new_channel})-(\ref{new_trace_constraint}) in an alternative
form. To this end, we note that the last $(r_1+p-k)$ entries of
$\tilde{\bby}_1$ come from only the noise. Since the noise is
white, see (\ref{noise_still_white}), we can omit these last
$r_1+p-k$ entries of $\tilde{\bby}_1$ without loss of generality.
Furthermore, we define
\begin{align}
\tilde{h}_{1\ell}=\Sigma_{1,\ell\ell},\quad 1\leq \ell\leq k-p
\label{alternative_channel_1}
\end{align}
Similarly, the first $r_2-p-s$ entries of $\tilde{\bby}_2$ come
from only the noise. Since the noise is white, see
(\ref{noise_still_white}), we can again omit these first $r_2-p-s$
entries of $\tilde{\bby}_2$ without loss of generality. Similarly,
we also define
\begin{align}
\tilde{h}_{2\ell}=\Sigma_{2,(r_2-k+\ell)\ell},\quad k-p-s+1\leq
\ell \leq k \label{alternative_channel_2}
\end{align}
Using the definitions in
(\ref{alternative_channel_1})-(\ref{alternative_channel_2}) and
omitting the entries of $\tilde{\bby}_1,\tilde{\bby}_2$ which
contain only noise, the channel in (\ref{new_channel}) can be
expressed as
\begin{align}
\tilde{Y}_{1\ell}&=\tilde{h}_{1\ell}\tilde{X}_\ell+\tilde{N}_{1\ell},\quad
\ell=1,\ldots,|\mathcal{S}_1|+|\mathcal{S}_c|\label{alternative_channel_3}\\
\tilde{Y}_{2\ell}&=\tilde{h}_{2\ell}\tilde{X}_\ell+\tilde{N}_{2\ell},\quad
\ell=|\mathcal{S}_1|+1,\ldots,|\mathcal{S}_1|+|\mathcal{S}_c|+|\mathcal{S}_2|
\label{alternative_channel_4}
\end{align}
where we used the definitions of
$\mathcal{S}_1,\mathcal{S}_c,\mathcal{S}_2$ given in
(\ref{set_1})-(\ref{set_2}) in conjunction with
(\ref{alternative_channel_1})-(\ref{alternative_channel_2}), and
$\{\tilde{N}_{1,\ell}\}_{\ell=1}^{\ell=|\mathcal{S}_1|+|\mathcal{S}_c|},\{\tilde{N}_{2,\ell}\}_{\ell=|\mathcal{S}_1|+1}^{|\mathcal{S}_1|+|\mathcal{S}_c|+|\mathcal{S}_2|}$
are i.i.d. Gaussian random variables with unit variance. The power
constraint on the channel input in (\ref{new_trace_constraint})
can be rewritten as
\begin{align}
\sum_{\ell=1}^{|\mathcal{S}_1|+|\mathcal{S}_c|+|\mathcal{S}_2|}E\left[\tilde{X}_{\ell}^2\right]\leq
\zeta P \label{alternative_constraint}
\end{align}
We note that the channel defined by
(\ref{alternative_channel_3})-(\ref{alternative_channel_4}) is a
parallel Gaussian broadcast channel with unmatched sub-channels,
whose capacity region is obtained in \cite{El_Gamal_Product}. In
particular, the capacity region of this channel can be obtained by
evaluating the following region
\begin{align}
R_0&\leq
\sum_{\ell\in\mathcal{S}_1\cup\mathcal{S}_c}I(U_{\ell};\tilde{Y}_{1\ell})\label{first_bound}\\
R_0&\leq
\sum_{\ell\in\mathcal{S}_2\cup\mathcal{S}_c}I(U_{\ell};\tilde{Y}_{2\ell})
\label{second_bound}\\
R_0+R_1&\leq
\sum_{\ell\in\mathcal{S}_{c2}}I(U_{\ell};\tilde{Y}_{1\ell})+\sum_{\ell
\in\mathcal{S}_{1}\cup\mathcal{S}_{c1}} I(X_\ell;\tilde{Y}_{1\ell})\label{third_bound} \\
R_0+R_2&\leq
\sum_{\ell\in\mathcal{S}_{c1}}I(U_{\ell};\tilde{Y}_{2\ell})+\sum_{\ell
\in\mathcal{S}_{2}\cup\mathcal{S}_{c2}} I(X_\ell;\tilde{Y}_{2\ell})\label{fourth_bound} \\
R_0+R_1+R_2&\leq
\sum_{\ell\in\mathcal{S}_{c2}}I(U_{\ell};\tilde{Y}_{1\ell})+\sum_{\ell\in\mathcal{S}_2\cup\mathcal{S}_{c2}}I(X_{\ell};\tilde{Y}_{2\ell}|U_{\ell})+\sum_{\ell
\in\mathcal{S}_{1}\cup\mathcal{S}_{c1}} I(X_\ell;\tilde{Y}_{1\ell})\label{fifth_bound} \\
R_0+R_1+R_2&\leq
\sum_{\ell\in\mathcal{S}_{c1}}I(U_{\ell};\tilde{Y}_{2\ell})+\sum_{\ell\in\mathcal{S}_1\cup\mathcal{S}_{c1}}I(X_{\ell};\tilde{Y}_{1\ell}|U_{\ell})+\sum_{\ell
\in\mathcal{S}_{2}\cup\mathcal{S}_{c2}}
I(X_\ell;\tilde{Y}_{2\ell})\label{sixth_bound}
\end{align}
by jointly Gaussian
$\{(U_{\ell},X_{\ell})\}_{\ell=1}^{|\mathcal{S}_1|+|\mathcal{S}_c|+|\mathcal{S}_2|}$~\cite{El_Gamal_Product},
where $\mathcal{S}_{c1}$ and $\mathcal{S}_{c2}$ are given by
\begin{align}
\mathcal{S}_{c1}=\left\{\ell\in\mathcal{S}_{c}:\tilde{h}_{1\ell}^2\geq
\tilde{h}_{2\ell}^2\right\}\\
\mathcal{S}_{c2}=\left\{\ell\in\mathcal{S}_{c}:\tilde{h}_{2\ell}^2\geq
\tilde{h}_{1\ell}^2\right\}
\end{align}
Hence, the capacity region of the channel in
(\ref{alternative_channel_3})-(\ref{alternative_channel_4}) is
given by the union of the rate triples $(R_0,R_1,R_2)$ satisfying
\begin{align}
R_0&\leq
\sum_{\ell\in\mathcal{S}_1\cup\mathcal{S}_{c}}C\left(\frac{\tilde{h}_{1\ell}^2\gamma_\ell
P_\ell}{1+\tilde{h}_{1\ell}^2\bar{\gamma}_\ell P_\ell}\right)\label{first_bound_g}\\
R_0&\leq
\sum_{\ell\in\mathcal{S}_2\cup\mathcal{S}_{c}}C\left(\frac{\tilde{h}_{2\ell}^2\gamma_\ell
P_\ell}{1+\tilde{h}_{2\ell}^2\bar{\gamma}_\ell P_\ell}\right)\label{second_bound_g}\\
R_0+R_1&\leq
\sum_{\ell\in\mathcal{S}_{c2}}C\left(\frac{\tilde{h}_{1\ell}^2\gamma_\ell
P_\ell}{1+\tilde{h}_{1\ell}^2\bar{\gamma}_\ell P_\ell}\right)+
\sum_{\ell\in\mathcal{S}_1\cup\mathcal{S}_{c1}}C\left(\tilde{h}_{1\ell}^2P_\ell\right)
\label{third_bound_g}\\
R_0+R_2&\leq
\sum_{\ell\in\mathcal{S}_{c1}}C\left(\frac{\tilde{h}_{2\ell}^2\gamma_\ell
P_\ell}{1+\tilde{h}_{2\ell}^2\bar{\gamma}_\ell P_\ell}\right)+
\sum_{\ell\in\mathcal{S}_2\cup\mathcal{S}_{c2}}C\left(\tilde{h}_{2\ell}^2P_\ell\right)
\label{fourth_bound_g}\\
R_0+R_1+R_2&\leq
\sum_{\ell\in\mathcal{S}_{c2}}C\left(\frac{\tilde{h}_{1\ell}^2\gamma_\ell
P_\ell}{1+\tilde{h}_{1\ell}^2\bar{\gamma}_\ell
P_\ell}\right)+\sum_{\ell\in\mathcal{S}_2\cup\mathcal{S}_{c2}}C\left(\tilde{h}_{2\ell}^2\bar{\gamma}_\ell
P_\ell\right)+
\sum_{\ell\in\mathcal{S}_1\cup\mathcal{S}_{c1}}C\left(\tilde{h}_{1\ell}^2P_\ell\right)
\label{fifth_bound_g}\\
R_0+R_1+R_2&\leq
\sum_{\ell\in\mathcal{S}_{c1}}C\left(\frac{\tilde{h}_{2\ell}^2\gamma_\ell
P_\ell}{1+\tilde{h}_{2\ell}^2\bar{\gamma}_\ell
P_\ell}\right)+\sum_{\ell\in\mathcal{S}_1\cup\mathcal{S}_{c1}}C\left(\tilde{h}_{1\ell}^2\bar{\gamma}_\ell
P_\ell\right)+
\sum_{\ell\in\mathcal{S}_2\cup\mathcal{S}_{c2}}C\left(\tilde{h}_{2\ell}^2P_\ell\right)
\label{sixth_bound_g}
\end{align}
for some
$\gamma_{\ell}=1-\bar{\gamma}_\ell\in[0,1],~\ell=1,\ldots,|\mathcal{S}_1|+|\mathcal{S}_c|+|\mathcal{S}_2|$,
and
$\{P_\ell\}_{\ell=1}^{|\mathcal{S}_1|+|\mathcal{S}_c|+|\mathcal{S}_2|}$
such that
$\sum_{\ell=1}^{|\mathcal{S}_1|+|\mathcal{S}_c|+|\mathcal{S}_2|}P_{\ell}=\zeta
P$, where $C(x)=(1/2)\log (1+x)$.

We now obtain the DoF region of the channel in
(\ref{alternative_channel_3})-(\ref{alternative_channel_4}) by
using (\ref{first_bound_g})-(\ref{sixth_bound_g}), which will
serve as an outer bound for the DoF region of the original channel
in
(\ref{general_gaussian_mimo_1})-(\ref{general_gaussian_mimo_2}).
To this end, we define
\begin{align}
\eta_j=\lim_{P\rightarrow
\infty}\frac{\sum_{\ell\in\mathcal{S}_j}C\left(\frac{\tilde{h}_{j\ell}^2\gamma_\ell
P_\ell}{1+\tilde{h}_{j\ell}^2\bar{\gamma}_\ell
P_\ell}\right)}{\frac{1}{2}\log P},\quad j=1,2 \label{def_eta}
\end{align}
We define $\delta_{c1}$ as follows
\begin{align}
\delta_{c1}&=\lim_{P\rightarrow
\infty}\frac{\sum_{\ell\in\mathcal{S}_{c1}}C\left(\frac{\tilde{h}_{1\ell}^2\gamma_\ell
P_\ell}{1+\tilde{h}_{1\ell}^2\bar{\gamma}_\ell
P_\ell}\right)}{\frac{1}{2}\log P}\label{def_alpha_1}\\
&\geq \lim_{P\rightarrow
\infty}\frac{\sum_{\ell\in\mathcal{S}_{c1}}C\left(\frac{\tilde{h}_{2\ell}^2\gamma_\ell
P_\ell}{1+\tilde{h}_{2\ell}^2\bar{\gamma}_\ell
P_\ell}\right)}{\frac{1}{2}\log P} \label{def_alpha_1_implies}
\end{align}
where the inequality comes from the fact that
$\tilde{h}_{1\ell}^2\geq
\tilde{h}_{2\ell}^2,~\ell\in\mathcal{S}_{c1}$. Similarly, we
define $\delta_{c2}$ as follows
\begin{align}
\delta_{c2}&=\lim_{P\rightarrow
\infty}\frac{\sum_{\ell\in\mathcal{S}_{c2}}C\left(\frac{\tilde{h}_{2\ell}^2\gamma_\ell
P_\ell}{1+\tilde{h}_{2\ell}^2\bar{\gamma}_\ell
P_\ell}\right)}{\frac{1}{2}\log P}\label{def_alpha_2}\\
&\geq \lim_{P\rightarrow
\infty}\frac{\sum_{\ell\in\mathcal{S}_{c2}}C\left(\frac{\tilde{h}_{1\ell}^2\gamma_\ell
P_\ell}{1+\tilde{h}_{1\ell}^2\bar{\gamma}_\ell
P_\ell}\right)}{\frac{1}{2}\log P} \label{def_alpha_2_implies}
\end{align}
where the inequality comes from the fact that
$\tilde{h}_{2\ell}^2\geq
\tilde{h}_{1\ell}^2,~\ell\in\mathcal{S}_{c2}$. Using
(\ref{def_eta}), (\ref{def_alpha_1}),(\ref{def_alpha_2_implies})
and the bound in (\ref{first_bound_g}), we get
\begin{align}
d_0\leq \eta_1+\delta_{c1}+\delta_{c2} \label{dof_outer_1}
\end{align}
Similarly, using (\ref{def_eta}),
(\ref{def_alpha_1_implies}),(\ref{def_alpha_2}) and the bound in
(\ref{second_bound_g}), we get
\begin{align}
d_0\leq \eta_2+\delta_{c1}+\delta_{c2} \label{dof_outer_2}
\end{align}
Using (\ref{def_alpha_2_implies}) and the rate bound in
(\ref{third_bound_g}), we get
\begin{align}
d_0+d_1&\leq \delta_{c2}+|\mathcal{S}_{c1}|+|\mathcal{S}_1|
\label{dof_outer_3}
\end{align}
Similarly, using (\ref{def_alpha_1_implies}) and the rate bound in
(\ref{fourth_bound_g}), we get
\begin{align}
d_0+d_2&\leq \delta_{c1}+|\mathcal{S}_{c2}|+|\mathcal{S}_2|
\label{dof_outer_4}
\end{align}
We next consider the rate bounds in (\ref{fifth_bound_g}) and
(\ref{sixth_bound_g}) to obtain bounds for $d_0+d_1+d_2$. To this
end, we note that
\begin{align}
C\left(\tilde{h}_{j\ell}^2\bar{\gamma_\ell}P_\ell\right)=C\left(\tilde{h}_{j\ell}^2
P_\ell\right)-C\left(\frac{\tilde{h}_{j\ell}^2\gamma_\ell
P_\ell}{1+\tilde{h}_{j\ell}^2\bar{\gamma}_\ell P_\ell}\right)
\end{align}
Using this identity, we get
\begin{align}
\lim_{P\rightarrow
\infty}\frac{\sum_{\ell\in\mathcal{S}_2\cup\mathcal{S}_{c2}}C\left(\tilde{h}_{2\ell}^2\bar{\gamma_\ell}P_\ell\right)}{\frac{1}{2}\log
P}&=\lim_{P\rightarrow
\infty}\frac{\sum_{\ell\in\mathcal{S}_2\cup\mathcal{S}_{c2}}C\left(\tilde{h}_{2\ell}^2
P_\ell\right)}{\frac{1}{2}\log P}-\lim_{P\rightarrow
\infty}\frac{\sum_{\ell\in\mathcal{S}_2\cup\mathcal{S}_{c2}}C\left(\frac{\tilde{h}_{2\ell}^2\gamma_\ell
P_\ell}{1+\tilde{h}_{2\ell}^2\bar{\gamma}_\ell
P_\ell}\right)}{\frac{1}{2}\log P}\\
&\leq |\mathcal{S}_2|+|\mathcal{S}_{c2}|-\lim_{P\rightarrow
\infty}\frac{\sum_{\ell\in\mathcal{S}_2\cup\mathcal{S}_{c2}}C\left(\frac{\tilde{h}_{2\ell}^2\gamma_\ell
P_\ell}{1+\tilde{h}_{2\ell}^2\bar{\gamma}_\ell
P_\ell}\right)}{\frac{1}{2}\log P}\\
&=|\mathcal{S}_2|+|\mathcal{S}_{c2}|-\eta_2-\delta_{c2}\label{defs_imply}
\end{align}
where in (\ref{defs_imply}), we used the definitions of
$\eta_2,\delta_{c2}$ given in (\ref{def_eta}),
(\ref{def_alpha_2}), respectively. Using (\ref{fifth_bound_g}) and
(\ref{defs_imply}), we get
\begin{align}
d_0+d_1+d_2&\leq
\delta_{c2}+|\mathcal{S}_2|+|\mathcal{S}_{c2}|-\eta_2-\delta_{c2}+|\mathcal{S}_{c1}|+|\mathcal{S}_1|
\\
&=|\mathcal{S}_2|+|\mathcal{S}_{c2}|+|\mathcal{S}_{c1}|+|\mathcal{S}_1|-\eta_2
\label{dof_outer_5}
\end{align}
Similarly, using the rate bound in (\ref{sixth_bound_g}), we can
get the following
\begin{align}
d_0+d_1+d_2&\leq|\mathcal{S}_2|+|\mathcal{S}_{c2}|+|\mathcal{S}_{c1}|+|\mathcal{S}_1|-\eta_1
\label{dof_outer_6}
\end{align}
Thus, we have obtained the DoF region of the channel in
(\ref{alternative_channel_3})-(\ref{alternative_channel_4}),
which, by combining (\ref{dof_outer_1})-(\ref{dof_outer_4}),
(\ref{dof_outer_5})-(\ref{dof_outer_6}), can be expressed as the
union of the triples $(d_0,d_1,d_2)$ satisfying
\begin{align}
d_0&\leq \min\{\eta_1,\eta_2\}+\delta_{c1}+\delta_{c2}\label{dof_outer_7}\\
d_0+d_1&\leq \delta_{c2}+|\mathcal{S}_{c1}|+|\mathcal{S}_1|\label{dof_outer_8}\\
d_0+d_2&\leq \delta_{c1}+|\mathcal{S}_{c2}|+|\mathcal{S}_2|\label{dof_outer_9}\\
d_0+d_1+d_2&\leq
|\mathcal{S}_1|+|\mathcal{S}_2|+|\mathcal{S}_{c1}|+|\mathcal{S}_{c2}|-\max\{\eta_1,\eta_2\}
\label{dof_outer_10}
\end{align}
for some non-negative $\eta_1,\eta_2,\delta_{c1},\delta_{c2}$ such
that $\eta_j\leq |\mathcal{S}_j|$, $\delta_{cj}\leq
|\mathcal{S}_{cj}|$. We define $\eta=\min\{\eta_1,\eta_2\}$. Using
this definition, we can enlarge the region in
(\ref{dof_outer_7})-(\ref{dof_outer_10}) as follows
\begin{align}
d_0&\leq \eta+\delta_{c1}+\delta_{c2}\label{dof_outer_11}\\
d_0+d_1&\leq \delta_{c2}+|\mathcal{S}_{c1}|+|\mathcal{S}_1|\label{dof_outer_12}\\
d_0+d_2&\leq \delta_{c1}+|\mathcal{S}_{c2}|+|\mathcal{S}_2|\label{dof_outer_13}\\
d_0+d_1+d_2&\leq
|\mathcal{S}_1|+|\mathcal{S}_2|+|\mathcal{S}_{c1}|+|\mathcal{S}_{c2}|-\eta
\label{dof_outer_14}
\end{align}
where $0\leq \eta\leq \min\{|\mathcal{S}_{1}|,|\mathcal{S}_2|\}$,
$\delta_{cj}\leq|\mathcal{S}_{cj}|$. Furthermore, we let
$\delta=\delta_{c1}+\delta_{c2}$ and define the region
$\mathcal{D}^{\rm out}$ as the union of the DoF triples
$(d_0,d_1,d_2)$ satisfying
\begin{align}
d_0&\leq \eta+\delta\label{dof_outer final_1}\\
d_0+d_1&\leq |\mathcal{S}_{c}|+|\mathcal{S}_1|\label{dof_outer_final 2}\\
d_0+d_2&\leq |\mathcal{S}_{c}|+|\mathcal{S}_2|\label{dof_outer_final 3}\\
d_0+d_1+d_2&\leq
|\mathcal{S}_1|+|\mathcal{S}_2|+|\mathcal{S}_{c}|-\eta
\label{dof_outer_final 4}
\end{align}
for some non-negative $\eta,\delta$ such that $\eta\leq
\min\{|\mathcal{S}_1|,|\mathcal{S}_2|\}$, and $\delta\leq
|\mathcal{S}_c|$. It is clear that $\mathcal{D}^{\rm out}$
contains the region in (\ref{dof_outer_11})-(\ref{dof_outer_14}),
and hence, is an outer bound for the DoF region of the Gaussian
MIMO broadcast channel with common and private messages given in
(\ref{general_gaussian_mimo_1})-(\ref{general_gaussian_mimo_2})-(\ref{trace_constraint}).

\section{Inner Bound}
\label{sec:inner_bound}

In this section, we provide an inner bound for the DoF region of
the Gaussian MIMO broadcast channel with common and private
messages, i.e., we show the achievability of the DoF region given
in Theorem~\ref{theorem_dof}. In particular, we provide two
different achievable schemes for the DoF region in
Theorem~\ref{theorem_dof}. The first one, presented in
Section~\ref{sec:inner_bound_dpc}, uses the DPC region in
Theorem~\ref{proposition_ach} directly. The second one, presented
in Section~\ref{sec:inner_bound_zf}, can be viewed as a variation
of the ZF scheme that eliminates the inter-user interference and
inter-transmit antenna interference by means of linear
pre-processing at the transmitter and linear post-processing at
the receivers.

\subsection{DPC-based Achievable Scheme}
\label{sec:inner_bound_dpc} We now obtain an inner bound for the
DoF region of the Gaussian MIMO broadcast channel with common and
private messages in
(\ref{general_gaussian_mimo_1})-(\ref{general_gaussian_mimo_2})-(\ref{trace_constraint})
by using the achievable scheme given in
Theorem~\ref{proposition_ach}. In particular, we make explicit
selections for the covariance matrices $\bbk_0,\bbk_1,\bbk_2$
involved in the achievable scheme of
Theorem~\ref{proposition_ach}, and show that the corresponding DoF
region is equal to the one given in Theorem~\ref{theorem_dof}. To
this end, we define the covariance matrices $\bbk_u$ as follows
\begin{align}
\bbk_u=(\xi P) \bbpsi_0 \left[\begin{array}{ccc} \bbomega \\
\bzero_{t-k\times k}
\end{array}
\right] \bblambda_u \left[~\bbomega^\top~~\bzero_{k\times
t-k}~\right] \bbpsi_0^\top,\quad u=0,1,2
\label{covariance_matrices}
\end{align}
where $\bblambda_u$ is a diagonal matrix of size $k\times k$.
$\xi$ in (\ref{covariance_matrices}) is selected to ensure that
${\rm tr}(\bbk_0+\bbk_1+\bbk_2)\leq P$. We next note the following
identity
\begin{align}
\frac{1}{\xi P}\bbh_j\bbk_u \bbh_j^\top &= \bbpsi_j\bbsigma_j
\left[~\bbomega^{-1}~~\bzero_{k\times t-k}\right]\bbpsi_0^{\top}
 \bbpsi_0 \left[\begin{array}{ccc} \bbomega \\
\bzero_{t-k\times k}
\end{array}
\right] \bblambda_u \left[~\bbomega^\top~~\bzero_{k\times
t-k}~\right] \bbpsi_0^\top \bbh_j^\top \label{gsvd_implies_1} \\
&=\bbpsi_j\bbsigma_j  \bblambda_u
\left[~\bbomega^\top~~\bzero_{k\times t-k}~\right] \bbpsi_0^\top
\bbh_j^\top \\
&=\bbpsi_j\bbsigma_j  \bblambda_u
\left[~\bbomega^\top~~\bzero_{k\times t-k}~\right] \bbpsi_0^\top
\bbpsi_0 \left[
\begin{array}{ccc}
\bbomega^{-\top}\\
\bzero_{t-k\times k}
\end{array}
\right] \bbsigma_j^\top\bbpsi_j^\top \label{gsvd_implies_2}\\
&=\bbpsi_j\bbsigma_j  \bblambda_u
\bbsigma_j^\top\bbpsi_j^\top,\quad j=1,2,~u=0,1,2
\end{align}
where (\ref{gsvd_implies_1}) and (\ref{gsvd_implies_2}) come from
the following identity
\begin{align}
\bbh_j=\bbpsi_j\bbsigma_j\left[~\bbomega^{-1}~~\bzero_{k\times
t-k}~\right] \bbpsi_0^\top
\end{align}
which is a consequence of the GSVD in
(\ref{decompose_h1})-(\ref{decompose_h2}). Thus, using the
covariance matrices $\bbk_0,\bbk_1,\bbk_2$ defined by
(\ref{covariance_matrices}) for the achievable scheme in
Theorem~\ref{proposition_ach}, we can get the following achievable
rates
\begin{align}
R_0(\bbk_0,\bbk_1,\bbk_2)&=\min_{j=1,2}\frac{1}{2}\log\frac{|(\xi
P)\bbpsi_j\bbsigma_j(\bblambda_0+\bblambda_1+\bblambda_2)\bbsigma_j^\top\bbpsi_j^\top+\bbi|}{|(\xi
P)\bbpsi_j\bbsigma_j(\bblambda_1+\bblambda_2)\bbsigma_j^\top\bbpsi_j^\top+\bbi|}
\\
R_1(\bbk_1,\bbk_2)&=\frac{1}{2}\log\frac{|(\xi
P)\bbpsi_1\bbsigma_1(\bblambda_1+\bblambda_2)\bbsigma_1^\top\bbpsi_1^\top+\bbi|}{|(\xi
P)\bbpsi_1\bbsigma_1\bblambda_2\bbsigma_1^\top\bbpsi_1^\top+\bbi|}\\
R_2(\bbk_2)&=\frac{1}{2}\log |(\xi
P)\bbpsi_2\bbsigma_2\bblambda_2\bbsigma_2^\top\bbpsi_2^\top+\bbi|
\end{align}
Using Slyvester's determinant theorem, i.e., $|\bba_{m\times
n}\bbb_{n\times m}+\bbi_{m\times m}|=|\bbb_{n\times
m}\bba_{m\times n}+\bbi_{n\times n}|$, these rates can be
expressed as follows
\begin{align}
R_0(\bbk_0,\bbk_1,\bbk_2)&=\min_{j=1,2}\frac{1}{2}\log\frac{|(\xi
P)(\bblambda_0+\bblambda_1+\bblambda_2)\bbsigma_j^\top\bbsigma_j+\bbi|}{|(\xi
P)(\bblambda_1+\bblambda_2)\bbsigma_j^\top\bbsigma_j+\bbi|}
\\
R_1(\bbk_1,\bbk_2)&=\frac{1}{2}\log\frac{|(\xi
P)(\bblambda_1+\bblambda_2)\bbsigma_1^\top\bbsigma_1+\bbi|}{|(\xi
P)\bblambda_2\bbsigma_1^\top\bbsigma_1+\bbi|}\\
R_2(\bbk_2)&=\frac{1}{2}\log |(\xi
P)\bblambda_2\bbsigma_2^\top\bbsigma_2+\bbi|
\end{align}
We note that $\bbsigma_1^\top\bbsigma_1,\bbsigma_2^\top\bbsigma_2$
are $k\times k$
($k=|\mathcal{S}_1|+|\mathcal{S}_c|+|\mathcal{S}_2|$) diagonal
matrices with the following structures
\begin{align}
(\bbsigma_1^\top\bbsigma_1)_{\ell\ell}& \left\{
\begin{array}{rcl}
&\hspace{-0.5cm}>0 &~~ {\rm if~~} 1\leq \ell \leq |\mathcal{S}_1|+|\mathcal{S}_c|\\
&\hspace{-0.5cm}=0 &~~ {\rm if~~}
|\mathcal{S}_1|+|\mathcal{S}_c|+1\leq \ell \leq
|\mathcal{S}_1|+|\mathcal{S}_c|+|\mathcal{S}_2|
\end{array}
\right.\label{structure_of_the_channel_matrix_1}\\
(\bbsigma_2^\top\bbsigma_2)_{\ell\ell}& \left\{
\begin{array}{rcl}
&\hspace{-0.5cm}=0 &~~ {\rm if~~} 1\leq \ell \leq |\mathcal{S}_1|\\
&\hspace{-0.5cm}>0 &~~ {\rm if~~} |\mathcal{S}_1|+1\leq \ell \leq
|\mathcal{S}_1|+|\mathcal{S}_c|+|\mathcal{S}_2|
\end{array}
\right. \label{structure_of_the_channel_matrix_2}
\end{align}
We next specify the diagonal matrices
$\bblambda_0,\bblambda_1,\bblambda_2$ as follows
\begin{align}
\Lambda_{0,\ell\ell}
&=
\left\{
\begin{array}{rcl}
&\hspace{-0.5cm}1 &~~ {\rm if~~} 1\leq \ell \leq \beta \\
&\hspace{-0.5cm} 0 &~~ {\rm if~~} \beta+1\leq \ell \leq
|\mathcal{S}_1|+\alpha_1 \\
&\hspace{-0.5cm} 1 &~~ {\rm if~~} |\mathcal{S}_1|+\alpha_1+1\leq
\ell \leq
|\mathcal{S}_1|+|\mathcal{S}_c|-\alpha_2 \\
&\hspace{-0.5cm} 0 &~~ {\rm if~~}
|\mathcal{S}_1|+|\mathcal{S}_c|-\alpha_2+1\leq
\ell \leq |\mathcal{S}_1|+|\mathcal{S}_c|+|\mathcal{S}_2|-\beta \\
&\hspace{-0.5cm} 1 &~~ {\rm if~~}
|\mathcal{S}_1|+|\mathcal{S}_c|+|\mathcal{S}_2|-\beta+1\leq \ell
\leq |\mathcal{S}_1|+|\mathcal{S}_c|+|\mathcal{S}_2|
\end{array}
\right. \label{structure_of_the_covariance_matrix_1}
\end{align}
\begin{align}
\Lambda_{1,\ell\ell} &= \left\{
\begin{array}{rcl}
&\hspace{-0.5cm}0 &~~ {\rm if~~} 1\leq \ell \leq \beta \\
&\hspace{-0.5cm} 1 &~~ {\rm if~~} \beta+1\leq \ell \leq
|\mathcal{S}_1|+\alpha_1 \\
&\hspace{-0.5cm}0 &~~{\rm if~~} |\mathcal{S}_1|+\alpha_1+1\leq
\ell \leq |\mathcal{S}_1|+|\mathcal{S}_c|+|\mathcal{S}_2|
\end{array}
\right. \label{structure_of_the_covariance_matrix_2}
\end{align}
\begin{align}
\Lambda_{2,\ell\ell} &= \left\{
\begin{array}{rcl}
&\hspace{-0.5cm}0 &~~ {\rm if~~} 1\leq \ell \leq |\mathcal{S}_1|+|\mathcal{S}_c|-\alpha_2 \\
&\hspace{-0.5cm} 1 &~~ {\rm if~~}
|\mathcal{S}_1|+|\mathcal{S}_c|-\alpha_2+1\leq \ell \leq
|\mathcal{S}_1|+|\mathcal{S}_c|+|\mathcal{S}_2|-\beta \\
&\hspace{-0.5cm}0 &~~{\rm if~~}
|\mathcal{S}_1|+|\mathcal{S}_c|+|\mathcal{S}_2|-\beta+1\leq \ell
\leq |\mathcal{S}_1|+|\mathcal{S}_c|+|\mathcal{S}_2|
\end{array}
\right. \label{structure_of_the_covariance_matrix_3}
\end{align}
where $0\leq \beta \leq \min\{|\mathcal{S}_1|,|\mathcal{S}_2|\}$,
$0\leq \alpha_j$, $\alpha_1+\alpha_2\leq |\mathcal{S}_c|$. These
selections of $\bblambda_0,\bblambda_1,\bblambda_2$ yield
\begin{align}
R_{01}(\bbk_0,\bbk_1,\bbk_2)&=\frac{1}{2} \log \frac{|(\xi
P)(\bblambda_0+\bblambda_1+\bblambda_2)\bbsigma_1^\top\bbsigma_1+\bbi|}{|(\xi
P)(\bblambda_1+\bblambda_2)\bbsigma_1^\top\bbsigma_1+\bbi|}\\
&=\frac{1}{2}\sum_{\ell=1}^{|\mathcal{S}_1|+|\mathcal{S}_c|}\log\frac{(\xi
P)(\Lambda_{0,\ell}+\Lambda_{1,\ell\ell}+\Lambda_{2,\ell\ell})(\bbsigma_1^\top\bbsigma_1)_{\ell\ell}+1}
{(\xi
P)(\Lambda_{1,\ell\ell}+\Lambda_{2,\ell\ell})(\bbsigma_1^\top\bbsigma_1)_{\ell\ell}+1}\label{structure_of_the_channel_matrix_1_implies}\\
&=\frac{1}{2}\sum_{\ell=1}^{\beta}\log \left((\xi
P)(\bbsigma_1^\top\bbsigma_1)_{\ell\ell}+1\right)
+\frac{1}{2}\sum_{\ell=|\mathcal{S}_1|+\alpha_1+1}^{|\mathcal{S}_1|+|\mathcal{S}_c|-\alpha_2}
\log \left((\xi P)(\bbsigma_1^\top\bbsigma_1)_{\ell\ell}+1\right)
\label{structure_of_the_covariance_matrix_implies}
\end{align}
where (\ref{structure_of_the_channel_matrix_1_implies}) comes from
the fact that
$\bblambda_0,\bblambda_1,\bblambda_2,\bbsigma_1^\top\bbsigma_1$
are diagonal by noting the structure of
$\bbsigma_1^\top\bbsigma_1$ stated in
(\ref{structure_of_the_channel_matrix_1}), and
(\ref{structure_of_the_covariance_matrix_implies}) is a
consequence of our $\bblambda_0,\bblambda_1,\bblambda_2$ choices
given in
(\ref{structure_of_the_covariance_matrix_1})-(\ref{structure_of_the_covariance_matrix_3}),
respectively. Equation
(\ref{structure_of_the_covariance_matrix_implies}) implies that
\begin{align}
\lim_{P\rightarrow \infty
}\frac{R_{01}(\bbk_0,\bbk_1,\bbk_2)}{\frac{1}{2}\log
P}=\beta+|\mathcal{S}_c|-\alpha_1-\alpha_2 \label{d0_inner_1}
\end{align}
Similarly, we have
\begin{align}
R_{02}(\bbk_0,\bbk_1,\bbk_2)&=\frac{1}{2} \log \frac{|(\xi
P)(\bblambda_0+\bblambda_1+\bblambda_2)\bbsigma_2^\top\bbsigma_2+\bbi|}{|(\xi
P)(\bblambda_1+\bblambda_2)\bbsigma_2^\top\bbsigma_2+\bbi|}\\
&=\frac{1}{2}\sum_{\ell=|\mathcal{S}_1|+1}^{|\mathcal{S}_1|+|\mathcal{S}_c|+|\mathcal{S}_2|}\log\frac{(\xi
P)(\Lambda_{0,\ell}+\Lambda_{1,\ell\ell}+\Lambda_{2,\ell\ell})(\bbsigma_2^\top\bbsigma_2)_{\ell\ell}+1}
{(\xi
P)(\Lambda_{1,\ell\ell}+\Lambda_{2,\ell\ell})(\bbsigma_2^\top\bbsigma_2)_{\ell\ell}+1}\label{structure_of_the_channel_matrix_2_implies}\\
&=\frac{1}{2}\sum_{\ell=|\mathcal{S}_1|+\alpha_1+1}^{|\mathcal{S}_1|+|\mathcal{S}_c|-\alpha_2}\log
\left((\xi
P)(\bbsigma_2^\top\bbsigma_2)_{\ell\ell}+1\right)\nonumber\\
&\quad
+\frac{1}{2}\sum_{\ell=|\mathcal{S}_1|+|\mathcal{S}_c|+|\mathcal{S}_2|-\beta+1}^{|\mathcal{S}_1|+|\mathcal{S}_c|+|\mathcal{S}_2|}
\log \left((\xi
P)(\bbsigma_2^\top\bbsigma_2)_{\ell\ell}+1\right)\label{structure_of_the_covariance_matrix_implies_1}
\end{align}
where (\ref{structure_of_the_channel_matrix_2_implies}) comes from
the fact that
$\bblambda_0,\bblambda_1,\bblambda_2,\bbsigma_2^\top\bbsigma_2$
are diagonal by noting the structure of
$\bbsigma_2^\top\bbsigma_2$ stated in
(\ref{structure_of_the_channel_matrix_2}), and
(\ref{structure_of_the_covariance_matrix_implies_1}) is a
consequence of our $\bblambda_0,\bblambda_1,\bblambda_2$ choices
given in
(\ref{structure_of_the_covariance_matrix_1})-(\ref{structure_of_the_covariance_matrix_3}),
respectively. Equation
(\ref{structure_of_the_covariance_matrix_implies_1}) implies that
\begin{align}
\lim_{P\rightarrow \infty
}\frac{R_{02}(\bbk_0,\bbk_1,\bbk_2)}{\frac{1}{2}\log
P}=\beta+|\mathcal{S}_c|-\alpha_1-\alpha_2 \label{d0_inner_2}
\end{align}
Hence, combining (\ref{d0_inner_1}) and (\ref{d0_inner_2}) yields
that
\begin{align}
d_0=\beta+|\mathcal{S}_c|-\alpha_1-\alpha_2
\end{align}
is an achievable DoF for the common message. We now consider the
first user's rate as follows
\begin{align}
R_{1}(\bbk_1,\bbk_2)&=\frac{1}{2} \log\frac{|(\xi
P)(\bblambda_1+\bblambda_2)(\bbsigma_1^\top\bbsigma_1)+\bbi|}{|(\xi
P)\bblambda_2
(\bbsigma_1^\top\bbsigma_1)+\bbi|}\\
&=\frac{1}{2} \sum_{\ell=1}^{|\mathcal{S}_1|+|\mathcal{S}_c|} \log
\frac{|(\xi
P)(\Lambda_{1,\ell\ell}+\Lambda_{2,\ell\ell})(\bbsigma_1^\top\bbsigma_1)_{\ell\ell}+1|}{|(\xi
P)\Lambda_{2,\ell\ell}(\bbsigma_1^\top\bbsigma_1)_{\ell\ell}+1|}\label{structure_of_the_channel_matrix_1_implies_1}\\
&=\frac{1}{2} \sum_{\ell=\beta+1}^{|\mathcal{S}_1|+\alpha_1} \log
\left((\xi
P)(\bbsigma_1^\top\bbsigma_1)_{\ell\ell}+1\right)\label{structure_of_the_covariance_matrix_implies_2}
\end{align}
where (\ref{structure_of_the_channel_matrix_1_implies_1}) comes
from the fact that
$\bblambda_1,\bblambda_2,\bbsigma_1^\top\bbsigma_1$ are diagonal
by noting the structure of $\bbsigma_1^\top\bbsigma_1$ stated in
(\ref{structure_of_the_channel_matrix_1}), and
(\ref{structure_of_the_covariance_matrix_implies_2}) is a
consequence of our $\bblambda_1,\bblambda_2$ choices given in
(\ref{structure_of_the_covariance_matrix_2})-(\ref{structure_of_the_covariance_matrix_3}),
respectively. Equation
(\ref{structure_of_the_covariance_matrix_implies_2}) implies that
\begin{align}
d_1=\alpha_1+|\mathcal{S}_1|-\beta
\end{align}
is an achievable DoF for the first user's private message. We
finally consider the second user's rate as follows
\begin{align}
R_{2}(\bbk_2)&=\frac{1}{2} \log |(\xi
P)\bblambda_2(\bbsigma_2^\top\bbsigma_2)+\bbi|\\
&=\frac{1}{2}
\sum_{\ell=|\mathcal{S}_1|+|\mathcal{S}_c|-\alpha_2+1}^{|\mathcal{S}_1|+|\mathcal{S}_c|+|\mathcal{S}_2|-\beta}
\log \left((\xi
P)(\bbsigma_2^\top\bbsigma_2)_{\ell\ell}+1\right)\label{structure_of_the_covariance_matrix_implies_3}
\end{align}
where (\ref{structure_of_the_covariance_matrix_implies_3}) comes
from (\ref{structure_of_the_channel_matrix_2}) and
(\ref{structure_of_the_covariance_matrix_3}). Equation
(\ref{structure_of_the_covariance_matrix_implies_3}) implies that
\begin{align}
d_2=\alpha_2+|\mathcal{S}_2|-\beta
\end{align}
is an achievable DoF for the second user's private message. Thus,
we have obtained an inner bound $\mathcal{D}^{\rm in}$ for the DoF
region of the Gaussian MIMO broadcast channel with common and
private messages, where $\mathcal{D}^{\rm in}$ consists of DoF
triples $(d_0,d_1,d_2)$ satisfying
\begin{align}
d_0&\leq |\mathcal{S}_c|-\alpha_1-\alpha_2+\beta\label{dof_inner_final_1}\\
d_1&\leq \alpha_1+|\mathcal{S}_1|-\beta \label{dof_inner_final_2}\\
d_2&\leq \alpha_2+|\mathcal{S}_2|-\beta \label{dof_inner_final_3}
\end{align}
for some non-negative $\alpha_1,\alpha_2,\beta$ such that
$\alpha_1+\alpha_2\leq |\mathcal{S}_c|,\beta\leq
\min\{|\mathcal{S}_1|,|\mathcal{S}_2|\}$.

As a final remark, we note that here we obtain an inner bound for
the DoF region of the Gaussian MIMO broadcast channel with common
and private messages without any recourse to the alternative
parallel channel-like representation of the Gaussian MIMO channel
given in (\ref{towards_a_new_channel}). Indeed, this inner bound
can also be obtained by using this alternative representation, and
in the next section, the ZF-based achievable scheme implicitly
uses this alternative scheme.

\subsection{ZF-based Achievable Scheme}
\label{sec:inner_bound_zf}

In this section, we provide an alternative achievable scheme to
show that the DoF region given in Theorem~\ref{theorem_dof} is
achievable. This alternative achievable scheme can be viewed as a
variation of the ZF scheme~\cite{zf_1,zf_2}, where the ZF scheme
is originally proposed for the Gaussian MIMO broadcast channel
with only private messages, i.e., without a common message. In
this ZF scheme, the transmitter eliminates the inter-user
interference via a linear pre-processing of its transmitted
signals. In particular, the transmitter sends each user's message
in the null space of the other user's channel gain matrix such
that each user sees an interference-free link between itself and
the transmitter. However, this complete elimination of the
inter-user interference can be accomplished only under certain
conditions on the ranks of the channel gain matrices
$\bbh_1,\bbh_2$, i.e., under certain conditions on the number of
transmit and receive antennas $t,r_1,r_2$. In
particular,~\cite{zf_1,zf_2} show that the ZF scheme can attain
the DoF for the private message sum rate\footnote{The DoF for the
private message sum rate is given by \begin{align} d^{\rm
sum}=\lim_{P\rightarrow \infty}\frac{R_1+R_2}{\frac{1}{2}\log
P}\nonumber \end{align}} when $r_1+r_2\leq t$. This restriction
comes from the fact that in the ZF scheme, each user's message is
sent through the null-space of the other user's channel gain
matrix. Alternatively, this restriction can be explained by
examining the methodology of the ZF scheme, which uses individual
singular value decompositions of the channel gain matrices
$\bbh_1,\bbh_2$ to obtain the pre-coding matrix of each
user~\cite{zf_1,zf_2}. However, by using the GSVD of the two
channel gain matrices simultaneously to obtain the precoding
matrices of the two users, this restriction can be removed as we
do here. In the variation of the ZF scheme we propose here, the
transmitter sends
\begin{align}
\bbx&= \bbpsi_0 \left[\begin{array}{ccc} \bbomega \\
\bzero_{t-k\times k}
\end{array}
\right]\left(\hat{\bbx}_1+\hat{\bbx}_c+\hat{\bbx}_2\right)
\label{zf_x}
\end{align}
where $\hat{\bbx}_1,\hat{\bbx}_c,\hat{\bbx}_2$ are given by
\begin{align}
\hat{\bbx}_1&=\left[~\hat{X}_{11}\ldots
\hat{X}_{1|\mathcal{S}_1|}~~\bzero_{1\times (|\mathcal{S}_c|+|\mathcal{S}_2|)}~\right]^\top\label{zf_x1}\\
\hat{\bbx}_c&=\left[~\bzero_{1\times
|\mathcal{S}_1|}~~\hat{X}_{c1}\ldots
\hat{X}_{c|\mathcal{S}_c|}~~\bzero_{1\times |\mathcal{S}_2|}~\right]^\top\label{zf_xc}\\
\hat{\bbx}_2&=\left[~\bzero_{1\times
(|\mathcal{S}_1|+|\mathcal{S}_c|)}~~\hat{X}_{21}\ldots
\hat{X}_{2|\mathcal{S}_2|}~\right]^\top \label{zf_x2}
\end{align}
Consequently, the received signal at the first user can be written
as
\begin{align}
\bby_1&=\bbh_1 \bbx+\bbn_1\\
&=\bbpsi_1\bbsigma_1\left[~\bbomega^{-1}~~\bzero_{k\times
t-k}~\right] \bbpsi_0^\top \bbpsi_0 \left[\begin{array}{ccc} \bbomega \\
\bzero_{t-k\times k}
\end{array}
\right]\left(\hat{\bbx}_1+\hat{\bbx}_c+\hat{\bbx}_2\right)+\bbn_1 \label{gsvd_implies_3}\\
&=\bbpsi_1\bbsigma_1
(\hat{\bbx}_1+\hat{\bbx}_c+\hat{\bbx}_2)+\bbn_1\\
&=\bbpsi_1\bbsigma_1 (\hat{\bbx}_1+\hat{\bbx}_c)+\bbn_1
\label{zf_implies}
\end{align}
where (\ref{gsvd_implies_3}) is a consequence of the GSVD and
(\ref{zf_x}), and (\ref{zf_implies}) comes from the fact that
$\bbsigma_1\hat{\bbx}_2=\bzero$. After multiplying $\bby_1$ by the
orthonormal matrix $\bbpsi_1^\top$, we get
\begin{align}
\hat{\bby}_1&=\bbpsi_1^\top\bby_1\\
&=\bbsigma_1 (\hat{\bbx}_1+\hat{\bbx}_c)+\hat{\bbn}_1
\end{align}
where $\hat{\bbn}_1=\bbpsi_1^\top\bbn_1$ is additive white
Gaussian noise with unit covariance matrix. Thus, the channel
outputs resulting from the use of the channel input defined by
(\ref{zf_x})-(\ref{zf_x2}) are given by
\begin{align}
\hat{\bby}_1&=\bbsigma_1 (\hat{\bbx}_1+\hat{\bbx}_c)+\hat{\bbn}_1
\label{zf_channel_1}
\\
\hat{\bby}_2&=\bbsigma_2 (\hat{\bbx}_c+\hat{\bbx}_2)+\hat{\bbn}_2
\label{zf_channel_2}
\end{align}
This equivalent form of the channel in
(\ref{zf_channel_1})-(\ref{zf_channel_2}), which results from the
use of the ZF scheme, imply that, since $\bbsigma_1$ and
$\bbsigma_2$ are diagonal, the ZF transforms the channel into a
parallel Gaussian broadcast channel with unmatched sub-channels,
where both users have access to $|\mathcal{S}_c|$ sub-channels
through which they observe a noisy version of $\hat{\bbx}_c$. In
addition to these common sub-channels, the $j$th user has access
to $|\mathcal{S}_j|$ sub-channels through which it observes a
noisy version of $\hat{\bbx}_j$, and the other user cannot observe
these sub-channels. Now, we consider independent Gaussian coding
across all sub-channels to obtain the DoF region given in
Theorem~\ref{theorem_dof}. In particular, we send the common
message through $|\mathcal{S}_c|-\alpha_1-\alpha_2$ sub-channels
of the $|\mathcal{S}_c|$ common sub-channels that both users can
access and $\beta$ sub-channels of each user's private
sub-channels which cannot be observed by the other user. The $j$th
user's private message is transmitted through $\alpha_j$
sub-channels of the $|\mathcal{S}_c|$ common sub-channels in
addition to the $|\mathcal{S}_j|-\beta$ sub-channels of the $j$th
user's private sub-channels that cannot be observed by the other
user. Consequently, this coding scheme yields the following
achievable rate triples
\begin{align}
R_0& \approx \frac{|\mathcal{S}_c|-\alpha_1-\alpha_2+\beta }{2} \log P \label{zf_rates_1} \\
R_1 &\approx \frac{\alpha_1+|\mathcal{S}_1|-\beta}{2} \log P \label{zf_rates_2}\\
R_2 & \approx \frac{\alpha_2+|\mathcal{S}_2|-\beta}{2}\log P
\label{zf_rates_3}
\end{align}
for any non-negative $\alpha_1,\alpha_2,\beta$ satisfying
$\alpha_1+\alpha_2 \leq |\mathcal{S}_c|,\beta \leq
\min\{|\mathcal{S}_1|,|\mathcal{S}_2|\}$. The achievable rate
triples given by (\ref{zf_rates_1})-(\ref{zf_rates_3}) imply that
the DoF region that ZF scheme can attain is equal to the one that
DPC attains, i.e., $\mathcal{D}^{\rm in}$, where $\mathcal{D}^{\rm
in}$ is the DoF region given in Theorem~\ref{theorem_dof}.

\section{Equivalence of the Inner and Outer Bounds}
\label{sec:equivalence}

We now show that the inner bound $\mathcal{D}^{\rm in}$ for the
DoF region of the Gaussian MIMO broadcast channel with common and
private messages given in
(\ref{dof_inner_final_1})-(\ref{dof_inner_final_3}) is equal to
the outer bound $\mathcal{D}^{\rm out}$ for the DoF region of the
Gaussian MIMO broadcast channel with common and private messages
given in (\ref{dof_outer final_1})-(\ref{dof_outer_final 4}).
$\mathcal{D}^{\rm in}$ is defined by the following equations
\begin{align}
d_0&\leq |\mathcal{S}_c|-\alpha_1-\alpha_2+\beta \label{FM_1}\\
d_1&\leq \alpha_1+|\mathcal{S}_1|-\beta \label{FM_2}\\
d_2&\leq \alpha_2+|\mathcal{S}_2|-\beta \label{FM_3}\\
0&\leq \alpha_1\label{FM_4}\\
0&\leq \alpha_2 \label{FM_5}\\
\alpha_1+\alpha_2 &\leq |\mathcal{S}_c|\label{FM_6}\\
0&\leq \beta \leq
\min\{|\mathcal{S}_1|,|\mathcal{S}_2|\}\label{FM_7}
\end{align}
We define $\alpha=\alpha_1+\alpha_2$, using which in
(\ref{FM_1})-(\ref{FM_7}), we get
\begin{align}
d_0&\leq |\mathcal{S}_c|-\alpha+\beta \label{FM_1_1}\\
d_1&\leq \alpha-\alpha_2+|\mathcal{S}_1|-\beta \label{FM_2_1}\\
d_2&\leq \alpha_2+|\mathcal{S}_2|-\beta \label{FM_3_1}\\
0&\leq \alpha-\alpha_2\label{FM_4_1}\\
0&\leq \alpha_2 \label{FM_5_1}\\
\alpha &\leq |\mathcal{S}_c| \label{FM_6_1}\\
0&\leq \beta \leq \min\{|\mathcal{S}_1,\mathcal{S}_2|\}
\label{FM_7_1}
\end{align}
We can eliminate $\alpha_2$ from (\ref{FM_1_1})-(\ref{FM_7_1}) by
using Fourier-Motzkin elimination, which yields
\begin{align}
d_0&\leq |\mathcal{S}_c|-\alpha+\beta \label{FM_1_2}\\
d_1+d_2&\leq \alpha+|\mathcal{S}_1|+|\mathcal{S}_2|-2\beta \label{FM_2_2} \\
d_1&\leq \alpha+|\mathcal{S}_1|-\beta \label{FM_3_2}\\
d_2&\leq \alpha+|\mathcal{S}_2|-\beta \label{FM_4_2}\\
0&\leq \alpha \leq |\mathcal{S}_c|\label{FM_5_2}\\
0&\leq \beta \leq \min\{|\mathcal{S}_1|,|\mathcal{S}_2|\}
\label{FM_6_2}
\end{align}
We next note that if $(d_0,d_1,d_2)$ is an achievable DoF triple,
so is $(d_0-t_1-t_2,d_1+t_1,d_2+t_2)$ for any $(t_1,t_2)$ such
that $0\leq t_1,0\leq t_2,t_1+t_2\leq d_0$. We define
\begin{align}
d_0^\prime &=d_0-t_1-t_2\\
d_1^\prime &=d_1+t_1 \\
d_2^\prime &=d_2+t_2
\end{align}
using which, (\ref{FM_1_2})-(\ref{FM_6_2}) can be expressed as
\begin{align}
d_0^\prime+t_1+t_2&\leq |\mathcal{S}_c|-\alpha+\beta \label{FM_1_3}\\
d_1^\prime+d_2^\prime-t_1-t_2&\leq \alpha+|\mathcal{S}_1|+|\mathcal{S}_2|-2\beta \label{FM_2_3} \\
d_1^\prime-t_1&\leq \alpha+|\mathcal{S}_1|-\beta \label{FM_3_3}\\
d_2^\prime-t_2 &\leq \alpha+|\mathcal{S}_2|-\beta \label{FM_4_3}\\
0&\leq t_1 \label{FM_5_3}\\
0&\leq t_2 \label{FM_6_3}\\
0&\leq \alpha \leq |\mathcal{S}_c|\label{FM_7_3}\\
0&\leq \beta\leq \min\{|\mathcal{S}_1|,|\mathcal{S}_2|\}
\label{FM_8_3}
\end{align}
We can eliminate $t_1$ from (\ref{FM_1_3})-(\ref{FM_8_3}) by using
Fourier-Motzkin elimination, which yields
\begin{align}
d_0^\prime +d_1^\prime +d_2^\prime &\leq
|\mathcal{S}_c|+|\mathcal{S}_1|+|\mathcal{S}_2|-\beta \label{FM_1_4} \\
d_0^\prime+d_1^\prime +t_2&\leq |\mathcal{S}_1|+|\mathcal{S}_c| \label{FM_2_4} \\
d_2^\prime-t_2&\leq \alpha+|\mathcal{S}_2|-\beta \label{FM_3_4}\\
d_0^\prime+t_2&\leq |\mathcal{S}_c|-\alpha+\beta \label{FM_4_4}\\
0&\leq t_2 \label{FM_5_4}\\
0&\leq \alpha \leq |\mathcal{S}_c|\label{FM_6_4}\\
0&\leq \beta \leq \min\{|\mathcal{S}_1|,|\mathcal{S}_2|\}
\label{FM_7_4}
\end{align}
After eliminating $t_2$ from (\ref{FM_1_4})-(\ref{FM_7_4}), we get
\begin{align}
d_0^\prime +d_1^\prime +d_2^\prime &\leq
|\mathcal{S}_c|+|\mathcal{S}_1|+|\mathcal{S}_2|-\beta \label{FM_1_5} \\
d_0^\prime+d_1^\prime+d_2^\prime &\leq
|\mathcal{S}_1|+|\mathcal{S}_c|+|\mathcal{S}_2|-\beta+\alpha
\label{FM_2_5}\\
d_0^\prime+d_1^\prime &\leq |\mathcal{S}_1|+|\mathcal{S}_c| \label{FM_2_6} \\
d_0^\prime+d_2^\prime&\leq |\mathcal{S}_2|+|\mathcal{S}_c| \label{FM_3_5}\\
d_0^\prime&\leq |\mathcal{S}_c|-\alpha+\beta \label{FM_4_5}\\
0&\leq \alpha \leq |\mathcal{S}_c|\label{FM_5_5}\\
0&\leq \beta \leq \min\{|\mathcal{S}_1|,|\mathcal{S}_2|\}
\label{FM_6_5}
\end{align}
We note that the bound in (\ref{FM_2_5}) is redundant. Since the
region described by (\ref{FM_1_5})-(\ref{FM_6_5}) is equal to the
region $\mathcal{D}^{\rm out}$ in (\ref{dof_outer
final_1})-(\ref{dof_outer_final 4}), this completes the proof.

\section{Conclusions}
In this work, we consider the Gaussian MIMO broadcast channel with
common and private messages and obtain the DoF region of this
channel. The crucial step in obtaining this result is to construct
a parallel Gaussian broadcast channel with unmatched sub-channels
from the Gaussian MIMO broadcast channel by using the GSVD. The
capacity region of the constructed parallel channel provides an
outer bound for the capacity region of the Gaussian MIMO broadcast
channel. Using the capacity result for the parallel channel, we
obtain an outer bound for the DoF region of the Gaussian MIMO
broadcast channel. We show that this outer bound can be attained
by the achievable scheme that combines a classical Gaussian coding
for the common message and DPC for the private messages. In
addition to the DPC scheme, we also show that a variation of the
ZF scheme can attain the DoF region of the Gaussian MIMO broadcast
channel with common and private messages.

\bibliographystyle{unsrt}
\bibliography{IEEEabrv,references2}

\begin{thebibliography}{10}

\bibitem{Shamai_MIMO}
H.~Weingarten, Y.~Steinberg, and S.~Shamai (Shitz).
\newblock The capacity region of the {G}aussian multiple-input multiple-output
  broadcast channel.
\newblock {\em {IEEE} Trans. Inf. Theory}, 52(9):3936--3964, Sep. 2006.

\bibitem{Hannan_Common}
H.~Weingarten, Y.~Steinberg, and S.~Shamai (Shitz).
\newblock On the capacity region of the multi-antenna broadcast channel with
  common messages.
\newblock In {\em IEEE ISIT}, Jul. 2006.

\bibitem{hannan_thesis}
H.~Weingarten.
\newblock {\em Multiple-input multiple-output broadcast systems}.
\newblock PhD thesis, Technion, Haifa, Israel, 2007.

\bibitem{Goldsmith_common}
N.~Jindal and A.~Goldsmith.
\newblock Optimal power allocation for parallel broadcast channels with
  independent and common information.
\newblock In {\em IEEE Intl. Symp. Inf. Theory}, page 215, Jun. 2004.

\bibitem{El_Gamal_Product}
A.~El Gamal.
\newblock Capacity of the product and sum of two unmatched broadcast channels.
\newblock {\em Problems of Information Transmission}, 16(1):3--23, Jan. 1980.

\bibitem{MIMO_Common_Private}
E.~Ekrem and S.~Ulukus.
\newblock On {G}aussian {MIMO} broadcast channels with common and private
  messages.
\newblock In {\em IEEE ISIT}, Jun. 2010.

\bibitem{MIMO_Common_Private_Shamai}
R.~Liu, T.~Liu, H.~V. Poor, and S.~Shamai.
\newblock {MIMO} {G}aussian broadcast channels with common messages.
\newblock In {\em Information Theory and Applications Workshop}, Jan. 2010.
\newblock Also available at
  http://ita.ucsd.edu/workshop/10/files/paper/paper\_1139.pdf.

\bibitem{GSVD}
C.~Paige and M.~A. Saunders.
\newblock Towards a generalized singular value decomposition.
\newblock {\em SIAM. J. Numer. Anal.}, 18(398-405), Jun. 1981.

\bibitem{zf_1}
L.-U. Choi and R.~D. Murch.
\newblock A transmit preprocessing technique for multiuser {MIMO} systems using
  a decomposition approach.
\newblock {\em IEEE Trans. Wireless Commun}, 3(1):20--24, Jan. 2004.

\bibitem{zf_2}
Q.~H. Spencer, A.~L. Swindlehurst, and M.~Haardt.
\newblock Zero-forcing methods for downlink spatial multiplexing in multiuser
  {MIMO} channels.
\newblock {\em IEEE Trans. Signal Process.}, 52(2):461--471, Feb. 2004.

\bibitem{Wornell}
A.~Khisti and G.~Wornell.
\newblock Secure transmission with multiple antennas {II}: The {MIMOME}
  channel.
\newblock {\em {IEEE} Trans. Inf. Theory}, 56(11), Nov. 2010.

\end{thebibliography}
\end{document}